\documentclass[pra,twocolumn,superscriptaddress]{revtex4-1}

\usepackage{amssymb}
\usepackage{graphicx}
\usepackage{dcolumn}
\usepackage{bm}
\usepackage{amsmath}
\usepackage{mathrsfs}

\usepackage{makecell}

\DeclareRobustCommand{\atan}{%
  \operatorname{atan}%
  \@ifnextchar2{_}{}%
}

\usepackage{collectbox}

\usepackage{textcomp}
\usepackage[unicode=true,bookmarks=true,bookmarksnumbered=false,bookmarksopen=false,breaklinks=false,pdfborder={0 0 1},backref=false,colorlinks=true]{hyperref}

\hypersetup{linkcolor=magenta,urlcolor=blue,citecolor=blue,pdfstartview={FitH},hyperfootnotes=false,unicode=true}

\def\be{\begin{equation}}
\def\ee{\end{equation}}
\def\bea{\begin{eqnarray}}
\def\eea{\end{eqnarray}}
\def\nn{\nonumber}

\def\ncsection#1{{\par{\vskip 5pt}\noindent\Large #1}}

\begin{document}
\title{Quantum Information Scrambling Through a High-Complexity Operator Mapping}
\author{Xiaopeng Li}
\affiliation{State Key Laboratory of Surface Physics, Institute of Nanoelectronics and Quantum Computing, and Department of Physics, Fudan University, Shanghai 200433, China}
\affiliation{Collaborative Innovation Center of Advanced Microstructures, Nanjing 210093, China}
\email{xiaopeng\_li@fudan.edu.cn}

%\author{Zhexuan Gong} 
%\affiliation{Department of Physics, Colorado School of Mines, Golden, Colorado 80401, USA} 

\author{Guanyu Zhu} 
\affiliation{Joint Quantum Institute, NIST/University of Maryland, College Park, Maryland 20742, USA} 

\author{Muxin Han}  
\affiliation{Department of Physics, Florida Atlantic University, 777 Glades Road, Boca Raton, FL 33431, USA}

\author{Xin Wang}
\affiliation{Department of Physics, City University of Hong Kong,
Tat Chee Avenue, Kowloon, Hong Kong SAR, China,
and City University of Hong Kong Shenzhen Research Institute, Shenzhen, Guangdong 518057, China} 

\begin{abstract}
Quantum information scrambling has attracted much attention amid the effort to reconcile the conflict between quantum-mechanical unitarity and the thermalizaiton-irreversibility in many-body systems. Here we propose an unconventional mechanism to generate quantum information scrambling through a high-complexity mapping from logical to physical degrees-of-freedom that hides the logical information into non-separable many-body-correlations. Corresponding to this mapping, we develop an algorithm to efficiently sample a Slater-determinant wavefunction and compute all physical observables in dynamics with a polynomial cost in system-size. The system shows information scrambling in the quantum many-body Hilbert space characterized by the spreading of Hamming-distance. At late time, we find emergence of classical diffusion dynamics in this quantum many-body system. We establish that the operator-mapping enabled growth in out-of-time-order-correlator exhibits exponential-scrambling behavior. 
%Our sampling-algorithm   is expected to be applicable in accelerating quantum chemistry calculations and building classically-simulatable quantum-machine-learning architectures that exploit quantum-interference effects. 
The quantum information-hiding mapping approach may shed light on the understanding of fundamental connections among computational complexity, information scrambling and quantum thermalization.
%Recently, quantum information scrambling has attracted much attention  amid the effort to reconcile the fundamental conflict between quantum-mechanical unitarity and the irreversibility of thermalization in quantum many-body systems. Here we propose an unconventional mechanism to generate quantum information scrambling through a high-complexity mapping from logical to physical degrees of freedom that hides the logical information into non-separable many-body correlations. Corresponding to this mapping, we develop an algorithm to efficiently collapse a slater-determinant wavefunction and  compute all physical observables in dynamics with a polynomial cost in system-size. The system shows information scrambling in the quantum many-body Hilbert space characterized by the spreading of Hamming distance defined by a set of a natural orbital bases, which can also be calculated with a time polynomial in system-size. We establish that the operator-mapping enabled growth in the out-of-time-order-correlator still exhibits scrambling behavior. The algorithm we developed to sample slater-determinant is potentially useful in accelerating  quantum chemistry calculations and building classically-simulatable quantum machine learning architectures that exploit quantum interference effects. The quantum information-hiding mapping approach would shed light on the understanding of fundamental connections among computamodelingtional complexity, information scrambling and quantum thermalization. 
\end{abstract}

\date{\today}

%\pacs{67.85.-d, 03.75.Mn, 05.30.Jp, 05.30.Rt}

\maketitle

%\ncsection{Introduction}
\section{Introduction} 
Recent developments in  engineering synthetic quantum devices have achieved unprecedented controllability over a wide range of quantum degrees of freedom~\cite{2010Hammerer-lightatom,2010Duan-TrappedIon,2013Xiang-SCcircuits,2014Georgescu-quantumsimulation,2015Lewenstein-RPP,2016LiRPP,2017Wendin-SCqubits}. Theoretically, it has been realized that there is an intricate difference between few- and many- qubit systems---a few-qubit system undergoing unitary evolution is easily reversible by a quantum circuit, whereas the reverse for many qubits is in general very difficult or practically impossible due to quantum thermalization despite of unitarity~\cite{1991_Deutsch-PRA,1994_Srednicki-PRE,2014_Shaffer_Irreversibility,2016_Kaufman-Thermalization,2016Garttner-Scrambling,2017Du-Scrambling}. This conflict between unitarity and irreversibility with many qubits undermines our  fundamental understanding of quantum thermalization in a closed quantum system.  To reconcile the conflict, the eigenstate thermalization hypothesis has been formulated theoretically~\cite{1991_Deutsch-PRA,1994_Srednicki-PRE} and confirmed numerically~\cite{2008Rigol-Thermalization}. Nevertheless, the microscopic mechanism of the thermalization  remains illusive. Amid the fast-growing research interests on synthetic quantum systems for the purposes of quantum information processing and beyond, 
a thorough understanding of this problem is key for the field to advance forward~\cite{1998ZhuQcontrol,2016PoggiQcontrol}.
%  of further progress}~\cite{1998ZhuQcontrol,2016PoggiQcontrol}. 

The problem of many-body unitarity also emerges in modeling quantum effects of black holes~\cite{2013_Almheiri-JHEP}. The black hole information and the recent firewall paradox have received enormous research interests in the last few years~\cite{roberts2016lieb,Roberts2015,aleiner2016microscopic,gu2017local,chowdhury2017onset,patel2017quantum,von2018operator,nahum2017operator,2017Chen-SYK,2017Yao-SYK,xu2018accessing,2018arXiv180202633R}. 
One puzzle is that the thermal nature of black-hole evaporation process through Hawking radiation obtained from   semi-classical calculations is fundamentally inconsistent with a full quantum mechanical description. 

%The holographic approach leads to an intriguing result: the black hole corresponds to quantum chaos in the boundary holographic models \cite{Shenker:2013pqa}. 
%This is mostly studied using the  $0+1$ dimensional Sachdev-Ye-Kitaev (SYK) model~\cite{sachdev1993gapless,KitaevSYK} and  its variants owing to the exact-solvable nature in large-N limit~\cite{2018arXiv180202633R,gu2017local,2017Chen-SYK,2017Yao-SYK}.
%Researchers are now approaching a consensus that the computational complexity underlying the unitary dynamical evolution of a many-body system may play an central role in understanding quantum information scrambling~\cite{2018Susskind-Complexity,2018Swingle-Complex}. 
%although a solid theoretical framework for that is still lacking. 

In both contexts of quantum thermalization and black-hole information paradox, researchers are now approaching a consensus that understanding how quantum information scrambles is crucial to resolve the issues in quantum many-body unitarity~\cite{Shenker:2013pqa}. In connecting quantum microscopic degrees of freedom to large-scale classical (or semiclassical) physical processes, the interplay of quantum chaos~\cite{Shenker:2013pqa}, operator growth~\cite{2007_Hayden_JHEP,2018arXiv180202633R}, and computational complexity~\cite{2018Susskind-Complexity,2018Swingle-Complex,harlow2013quantum} is of deep fundamental interest. 
%expected to be  most interesting. 
%A full description of information scrambling requires modelling the interplay of quantum chaos, computational complexity, and a solid footing to connect microscopics to classical (or semiclassical) physical processes. 
%understanding how quantum information scrambles is crucial for   resolving the issues in quantum many-body unitarity. 
Recent developments following  the idea of the holographic anti-de Sitter/conformal field theory correspondence~\cite{Maldacena:1997re} 
%employing boundary holographic models to describe the bulk quantum black hole. 
have explored the quantum chaotic aspect  through the $0+1$ dimensional Sachdev-Ye-Kitaev (SYK) model~\cite{sachdev1993gapless,KitaevSYK} and its variants~\cite{2018arXiv180202633R,gu2017local,2017Chen-SYK,2017Yao-SYK}, whose out-of-time-order-correlators (OTOC) are analytically solvable at large-N limit.
% is the crucial microscopic ingredient underlying the conflicts rising in quantum many-body unitarity. 
However, other equally important aspects such as operator growth, computational complexity, and emergence of semi-classical descriptions 
%which are particularly important from quantum information perspectives 
are not well-captured by the previously studied analytically-solvable models~\cite{sachdev1993gapless,KitaevSYK,2018arXiv180202633R}, deserving novel theoretical ideas for further progress~\cite{2018Susskind-Complexity,2018Swingle-Complex}. 
%Recent studies suggest the computational  complexity to connect the microscopic degrees of freedom to physical observables in the scrambling process is one key ingredient so far not well understood~\cite{2018Susskind-Complexity,2018Swingle-Complex}. 
% The aspects of operator growth and computation complexity are not captured by previously studied analytic models~\cite{Stanford}. %and microscopic models or mechanism supporting such  aspects are lacking. operator growth and connecting quantum microscopic degrees of freedom to semi-classical description not  analytically captured by the SYK. 

Here we propose an unconventional approach for quantum information scrambling by applying a high complexity mapping. In our proposed mechanism, despite that the quantum state follows integrable unitary evolution, quantum information is gradually lost in a physical setting, morphing into non-separable many-body correlations because of the high complexity of the mapping between integrals of motion and physical observables in the system. This mechanism is demonstrated by a  specific carefully-designed operator mapping.  We show that the system exhibits information scrambling in the exponentially-sized Hilbert space. This dynamical process is quantified  by a Hamming distance~\cite{hauke2015many}. Corresponding to the mapping, we develop an algorithm and show the system is efficiently tractable on a classical computer---the time-cost scales polynomially in the system-size. 
At late-time quantum dynamics, we find a classical diffusion description emerges out of the microscopic quantum degrees of freedom. 
%which can be calculated with a polynomial time-cost for this model on a classical computer.  
%Despite the relatively low computational time cost  (polynomial) in calculating all equal-time observables,  we still find that the physical system develops quantum chaotic  growth in the out-of-time-order correlator (OTOC) ~\cite{larkin1969quasiclassical}.
%To connect to previous works, we also provide the results for out-of-time-order correlations~\cite{larkin1969quasiclassical} to reconfirm  information scrambling physics. 
The mapping-complexity in our proposed mechanism can be in principle generalized to more generic cases, which we expect to  open up a wide window to investigate the fundamental interplay of computational complexity, quantum thermalization and information scrambling. 
%A byproduct yet still worth  remarking is that the sampling-algorithm we developed is expected to be useful in accelerating quantum chemistry calculations and building classically-simulatable quantum-machine-learning architectures.

%Our constructed exact model would also shed light on the study of quantum Kolmogorov-Arnold-Moser theorem. 
%From the applicational perspective, our study provides a non-trivial way to hide quantum information through the unitary evolution circuit, and a decoding scheme to recover the information from the high-complexity map. 

%Corresponding to this mapping, we develop an algorithm to efficiently sample a slater-determinant wavefunction and compute all physical observables in dynamics with a polynomial cost in system-size. The system shows information scrambling in the quantum many-body Hilbert space characterized by the spreading of Hamming-distance. We establish that the operator-mapping enabled growth in out-of-time-order-correlator exhibits exponential-scrambling behavior. 

%\ncsection{Results}

\section{Quantum information scrambling through a high-complexity operator-mapping}
For a generic quantum system the time evolution of a physical observable ${\cal O}$ is described by 
the expectation value of the Heisenberg operator ${\cal O} (t) = e^{iHt}  {\cal O} e^{-iHt} $, with $H$ the system Hamiltonian. 
Considering a quantum state $|\Psi\rangle$ with information encoded by a set of few-body observables as ${\cal O}_\alpha |\Psi \rangle = \lambda_\alpha |\Psi\rangle$---for example these operators could be local pauli operators or stabilizers---its quantum evolution is deterministically given  by ${\cal O}_\alpha (-t) |\Psi (t) \rangle = \lambda_\alpha |\Psi (t)\rangle$. Through this dynamical process the information becomes hidden into ${\cal O}_\alpha (-t)$, which in general involve highly nonlocal many-body operators~\cite{2007_Hayden_JHEP,2008Susskind-Scramble,hastings2010locality,brown2010convergence,2018arXiv180202633R}. 
%Since the information is stored in ${\cal O}_{\alpha} (-t)$, we can define a logical basis as the eigenbaisis of ${\ca 
Tracking the mapping $\{ {\cal O}_\alpha \} \mapsto \{ {\cal O}_\alpha (-t)\}$ for a generic  Hamiltonian is exponentially complex, making the extraction of  quantum information after long-time evolution practically impossible, i.e., quantum information scrambles under the time evolution. The scrambling process can thus be treated as quantum information initiated in few-body degrees of freedom hidden into non-separable many-body correlations through a high-complexity mapping. 
Mathematically, we can define a logical basis that stores the quantum information according to the eigenbasis of ${\cal O}_{\alpha} (-t)$, and study the information scrambling in the physical degrees of freedom. 
The information scrambling  is thus basis-dependent and as a result the intrinsic physics soely relies on the mapping between logical and physical basis (see an illustration in TABLE~I). 
It is worth emphasizing here that this aspect is not captured by the well-studied SYK model~\cite{sachdev1993gapless,KitaevSYK} or its variants despite they are analytically solvable. 

It is evident that the capability of the mapping to build in non-separable many-body correlations is required to connect the information-preserving logical basis to the scrambling physical basis, 
%It is evident that the ${\cal O}_\alpha (t)$ operators having a non-separable many-body nature is necessary for information scrambling, 
but it is unclear whether the mapping having an exponential complexity is also necessary or not. Finding  a relatively lower-complexity mapping from few- to many-body operators that still causes information scrambling is not only of fundamental interest but would also assist in constructing exact models for  nontrivial quantum many-body dynamics. 
In constructing such a mapping, we require that 
(i) the mapping has to be complicated enough to enable scrambling,   
(ii) the mapping is numerically tractable with low-cost, or more specifically having polynomial computational complexity, and 
(iii) the locality of the quantum evolution must be respected by the mapping---the dynamics in both logical and physical basis should be local. 
We remark here that the Jordan Wigner~\cite{Lin_OTOC_PRB} or the Clifford gate unitary mapping~\cite{von2018operator} does not satisfy these requirements. 
In the following, we devise a suitable mapping based on spin-$1/2$ qubits. % or equivalently a spinless fermion model. 
We emphasize here that the proposed theoretical ideas are rather generalizable than restricted to the particular mapping.

\begin{table*}[hpt]
\label{map}
    %\begin{ruledtabular}
    \begin{tabular}
{p{0.14\textwidth} | p{0.25\textwidth} | p{0.25\textwidth}  } \hline \nonumber
    \,  			& \thead{Logical basis ($f$, $\tau$) } 		&\thead{ Physical basis  ($c$, $\sigma$)} 	\\ \hline
Information 	& Preserved 					& Scrambled (hidden) \\  \hline 
Basis-mapping &  \ldots, $ |0011001\rangle$, \ldots 	&\ldots, $|0 01 01 0 0 01 \rangle$, \ldots  \\    \hline 
Hilbertspace Dimension 
			&  $\left( \begin{array}{ll} L_\tau \\ N \end{array} \right)$
								&  $\left( \begin{array}{ll} L_\tau = L+1-N \\ N \end{array} \right)$ \\ \hline 
Hamiltonian 	&$H = \sum_k \left(f_k ^\dag f_{k+1} + H.c. \right)$	& $H = P\sum_j \left(c_j ^\dag c_{j+1} + H.c. \right) P$ \\ \hline 
Operator growth
			&$f_k^\dag f_{k'} $ remains quadratic	 		& $c_j^\dag  c_{j'} $ becomes scrambled \\ \hline 
    \end{tabular}
    %\end{ruledtabular}
    \caption{The operator-mapping approach. In our designed framework, quantum information is stored in logical basis but scrambled in the physical basis. In the mapping from the logical to physical basis (both of which are represented by binary bit strings), we replace `$0$' and `$1$' by `$0$' and `$01$', respectively, then remove `$0$' at the left-end to handle boundary effects.  The mapping is bijective, and the Hilbertspace dimensions of the two basis are identical.}
\end{table*}

Consider a one-dimensional spin chain, $\sigma_{j} ^{x,y,z}$, with the site index $j \in [1, L]$. We assume that  the total spin $S_z$  component is conserved, and shall work in the Hilbert space sector having $\sum_{j=1}^L \sigma^z_j /2 = S_z$. This $S_z$ conservation holds in the quantum dynamics to be specified below. 
We introduce a  highly nonlocal projective mapping as, 
\bea 
\textstyle \tau^z_1 &=& \textstyle \sigma^z_1 P \nn  \\
%\tau^z_{k \neq 1 }   &=& \sum_{S_z'=-\frac{k-1}{2} } ^ { {\rm min} (\frac{ k-1} {2}, S_z-1) } \delta_{S_z', \sum_{j'=1} ^{k-1} \tau^z _{j'} /2}  P\sigma^z_{k+S_z'} P \nn 
\textstyle \tau^z_{k \neq 1 }   &=& \textstyle \sum_{ \vec{s}_k} \sigma^z_{k+\sum_{j=1}^{k-1} (s_j + 1/2)} Q^{k}_{\vec {s} } P 
\label{eq:map} 
\eea 
where the two projectors are 
$P = \prod_{j=1}^{L-1} [1-(\sigma^z_j +1)(\sigma^z _{j+1} +1)/4] $, 
$Q_{\vec {s}}^k = \prod_{j=1}^{k-1} [s_j \tau_j ^z + 1/2]$, 
with $\vec{s}_k = \{ s_1, s_2, \ldots, s_{k-1}\}$, $s_j = \pm 1/2$, $k\in [1, L_\tau]$, ($L_\tau  \equiv L/2+1-S_z$). 
Correspondingly, we have 
$ 
\tau_k ^+ \tau_{k'} ^- 
= \sum_{\vec{s}_k} \sigma^+_{k + \sum_{j=1}^{k-1} (s_j + 1/2)} Q_{\vec{s} } ^k  \sum_{\vec{s}_{k'}'} \sigma^-_{k' + \sum_{j=1}^{k'-1} s_j ' + 1/2}
Q_{\vec{s}' } ^{k'}P, 
$
with $\sigma^{\pm}  = (\sigma^x \pm i\sigma_y)/2$. 
These $\tau$ operators  satisfy spin-$1/2$ algebra.  
%In the projected space, the operators $\tau_z$ take eigenvalues $\pm 1$, and the projected Hilbert space can be completely labeled by these eigenvalues. 
%The operators $\tau_x$ and  $\tau_y$ are defined correspondingly, and $\tau_{x,y,z}$ satisfy spin-1/2 algebra. 
Although the mapping appears a bit cryptic, it becomes more explicit when looking at the transformation of eigenbasis of $\tau_z$s and $\sigma_z$s. Taking an eigenstate of $\tau_z$s, say $|0011001\rangle$, we first map each `$0$' to `$0$', and each `$1$' to `$01$', and then remove the left-most `$0$'. The transformed state is an eigenstate of $\sigma_z$s, with the example state mapped to $|001010001\rangle$ (see TABLE~I).  This mapping was previously used to study the ground-state Luttinger liquid~\cite{gomez1993generalized,2009_Cheong_PRB} and also excited-state many-body localization~\cite{2017_Li_Bubble}, but an operator form of the mapping was not known, and an efficient algorithm to calculate physical observables under the mapping was technically lacking---the calculation in the previous literature was consequently restricted to small system-sizes~\cite{2009_Cheong_PRB}.

We shall consider a quantum state with information encoded as 
$\tau_k^z |\Psi \rangle = (2 m_k ^0-1)|\Psi\rangle$, that undergoes dynamical evolution  according to the Hamiltonian 
\be 
 H = \frac{1}{2} \sum_{k=1} ^{L_\tau} \left[ \tau ^x_k  \tau ^x_{k+1}  + \tau ^y _k \tau^y _{k+1} \right]. 
\label{eq:logicalHam} 
\ee 
An open boundary condition  is adopted in this work. We note here that the lattice constant is set to be the length unit and that the tunneling is the energy unit. 
In the $\sigma$ basis, the system is still local, and we have 
\be 
H = \frac{1}{2} \sum_{j=1} ^{L} P \left[ \sigma ^x_j  \sigma ^x_{j+1}  + \sigma ^y _j \sigma^y _{j+1} \right]P, 
\label{eq:physicalHam}
\ee 
and  $\sigma_j ^z |\Psi \rangle = (2 n_j^0 -1) |\Psi\rangle$.  The eigenvalues $m_k^0$ and $n_j^0$ are related by the transformation-rule specified above. 

The dynamics in this system is more transparent in the fermion picture through  a Jordan-Wigner transformation, where we introduce 
 $f_k$ and $c_j$ corresponding to $\tau$ and $\sigma$ degrees of freedom,  respectively. In the fermion picture, total $S_z$ conservation implies that 
the total particle number $N$ ($=S_z + L/2$) is conserved. 
%Taking  hard-core fermion model 
%\be 
%H = \sum_j \left( c_j ^\dag c_{j+1} + H.c.\right) + Vn_j n_{j+1 } 
%\ee 
%with the neighboring interaction $V$ being infinite, 
%the dynamics in the logical basis follows Eq.~\eqref{eq:logicalHam}. 
Corresponding to  Eq.~\eqref{eq:logicalHam}, $f_k$ operators follow free fermion dynamics with a  Hamiltonian 
$\sum_k ( f_k ^\dag f_{k+1} + H.c. )$, and $c_j$ follow dynamics of  a spinless hard-core fermion model with infinite nearest neighbor interaction, 
whose ground state is shown to be an interacting Luttinger liquid~\cite{gomez1993generalized,2009_Cheong_PRB}.  
To proceed, we introduce  the natural orbital bases $C_j (t) $ associated with $c$ operators by diagonalizing the time-dependent correlation function by  
\be
 \langle \Psi (t) | c_j ^\dag  c_{j'}  | \Psi (t )  \rangle = \sum_i \lambda_i (t)  U^c_{j, i} (t) U_{ j', i} ^{c*} (t),
\ee  
and $C_i  = \sum_j U^c_{j, i} c_j$. 
The natural orbital bases associated with $f$ operators are defined in the same way as $F_k$, which is given by $F_k (t)= e^{-iHt} f_k e^{iHt} $.  
During the dynamical evolution, the quantum information is thus stored as $F_k ^\dag F_k |\Psi ( t)\rangle = m_k ^0 |\Psi\rangle$, whereas in the $\sigma$ bases (or equivalently the $c$ bases)  
 the dynamical  $\sigma^z_j(-t)$ operator holding the information has a  non-separable many-body character. 
Under this mathematical construction, the degrees of freedom $\tau$ and $\sigma$ correspond to logical and physical basis, respectively. Properties of the mapping between the two are summarized in Table~I.  
%Mathematically, this dual feature allows us to track the quantum information with $\tau$s, and at the same time study the interesting many-body quantum dynamics in the $\sigma$s. In the following, we assume physical access to the $\sigma$ degrees of freedom,  and will thus refer to $\tau$ and $\sigma$  as logical and physical degrees of freedom, respectively. This dual feature is also expected to shed light on understanding operator growth in black hole scrambling physics~\cite{2018arXiv180202633R,nahum2017operator}. 

\section{Method and complexity for computing physical observables} 
Despite the many-body nature of the mapping from the logical to physical qubits in Eq.~\eqref{eq:map}, we find that the computational complexity to extract the physical observables is polynomial in the system size. 
%Starting from a product state in the logical spin $\tau$ basis, 
For the dynamical quantum state, 
$
|\Psi (t) \rangle =[ F_k ^{\dag} (t)]^{m_k ^0}  |0\rangle, 
$
%with $d_k  ^\dag (t) = e^{-iHt} d_k e^{iHt} $, and $m_{k}^0$ the occupation numbers fixed according to the initial state.  
the time-dependent wave function $\Psi_{\vec m}$ in the basis of $|\vec{m} \rangle =  \prod_k [ f_k^\dag (0) ]^{m_{k} } |0\rangle$ 
is a Slater-determinant. It is then straightforward to calculate  the few-body observables in terms of the logical qubits  via Wick theorem.  
However the Wick theorem does not carry over to  observables composed of physical qubits.   
A direct calculation of the expectation value of such a physical observable ${\cal O}$,  
$
\langle {\cal O} \rangle  = \sum_{\vec{m}, \vec{m}'} {\cal O}_{\vec{m} \vec{m}'} \Psi_{\vec m} ^* (t) \Psi_{\vec{m}'} (t), 
$
requires evaluating an exponentially large (in the system size) number of Slater-determinants, and is thus intractable for large systems. 

To overcome the exponential difficulty, we develop an algorithm to explicitly sample the Slater determinants efficiently with the computational omplexity reduced to polynomial. 
%Here, we provide an explicit algorithm to sample the Slater determinants which do not require calculation of exponentially large number of determinants. 
This algorithm is inspired by a recent work on boson sampling which proposes a sampling method for permanents~\cite{clifford2018classical}. 
Considering a Slater-determinant state represented by a matrix $U$ where each column represents a single-particle wave function. 
The algorithm contains three steps. The first step is to generate a random permutation of the integer sequence $[1,\ldots N]$, which has complexity $O(N)$. The generated sequence is denoted as a vector $\vec{v}$. The second step is to iteratively sample $x_k$ ($k$ is from $1$ to $N$ in the iteration) according to a conditional probability distribution 
$$
P (x_k| x_1,  \ldots x_{k-1} ; v_1, \ldots v_k ) \propto \frac{1}{k!} |{\rm Det} \left[U_{x_1 \ldots x_k, v_1 \ldots v_k }\right]|^2. 
$$  
%with ${\vec \eta} \backslash v_{k+1}, \ldots v_{N}$ meaning to keep the elements in ${\vec \eta}$ after  excluding $v_{k+1} \ldots v_{N}$.  
Then we choose an all zero vector $\vec {m}$ and then set $m_{x_k} =1 $. The vector $\vec {m}$ satisfies the required probability distribution $|\Psi_{\vec m} |^2$ because 
\be 
\frac{1}{N!}  \sum_{\vec v} \left[ \prod _k  P (x_k| x_1,  \ldots x_{k-1} ; v_1, \ldots v_k ) \right] =|\Psi_{\vec m} |^2, % |{\rm Det}\, U_{{\vec x}, {\vec n} } |^2, 
\ee 
which follows from 
\bea  
\sum_{x_k} |{\rm Det} \, U_{x_1 \ldots x_k; v_1 \ldots \nu_k} | ^2 %\nn \\  
 = \sum_{v \in [v_1 \ldots v_k]} | {\rm Det} \, U_{x_1 \ldots x_{k-1}; [v_1 \ldots v_k]\backslash v]} |^2, \nn
\eea
with $[v_1 \ldots v_k] \backslash v$ keeping the elements  in $[v_1 \ldots v_k]$ except $v$. 
The sampling algorithm substantially improves the calculation efficiency compared to the  previously used approach in studying ground state Luttinger liquids~\cite{2009_Cheong_PRB}, and allows to treat much larger system sizes. 
%This algorithm can also be adapted to other contexts such as determinant quantum Monte Carlo, where the sampling is performed through a Markov chain approach. Our algorithm allows explicit and independent sampling without any autocorrelation.  

%{\it Error and complexity analysis.} 
%In calculating physical observables,  we need efficient transformation between  the physical state basis $\prod_j [c_j ^\dag ]^{n_j} | 0\rangle$  and the logical basis $\prod_k [d_k ^\dag ]^{m_k} |0\rangle$, or equivalently we need a mapping between the vectors $\vec{n}$ and $\vec {m}$.  
The major cost to calculate the physical observables arises from  sampling the determinants. 
In the algorithm provided above, the complexity for $M_s$ samples scales as $O(M_s L_\tau N^{z+1} ) $, with $N^z$ the computational complexity to calculate the determinant of a $N\times N$ matrix ($z = 2.373$ with best known algorithm). Since the sampling error scales as $\epsilon \sim 1/\sqrt{M_s}$, the computational complexity to calculate one physical observable is  $O(\delta^{-2} L_\tau N^{z+1})$, with $\delta$ the required error threshold. 

The overall computation complexity to calculate one physical observable is reduced to  $O(L_\tau N^{3.373} /\delta^2)$ with $\delta $ the desired error threshold. The  dynamics for the interacting physical system defined by the mapping 
(Eq.~\eqref{eq:map}) is thus {\it almost} exactly solvable, in the sense that the computational complexity does not scale exponentially.

%Details of this algorithm are provided in Methods. 
%The overall computation complexity to calculate one physical observable is reduced to  $O(L_\tau N^{3.373} /\delta^2)$ with $\delta $ the desired error threshold. The  dynamics for the interacting physical system defined by the mapping  (Eq.~\eqref{eq:map}) is thus {\it almost} exactly solvable, in the sense that the computational complexity does not scale exponentially. 

We mention here that the matrix elements of the dynamical many-body operator 
$\langle \vec{m}  | e^{-iHt } c_j ^\dag c_j e^{iHt} | \vec{m}'\rangle$ holding the information in the physical basis, can be calculated at a time cost which scales polynomially, much more efficient than the exponential scaling in conventional settings.  

\begin{figure}[htp]
\includegraphics[width=.9\linewidth]{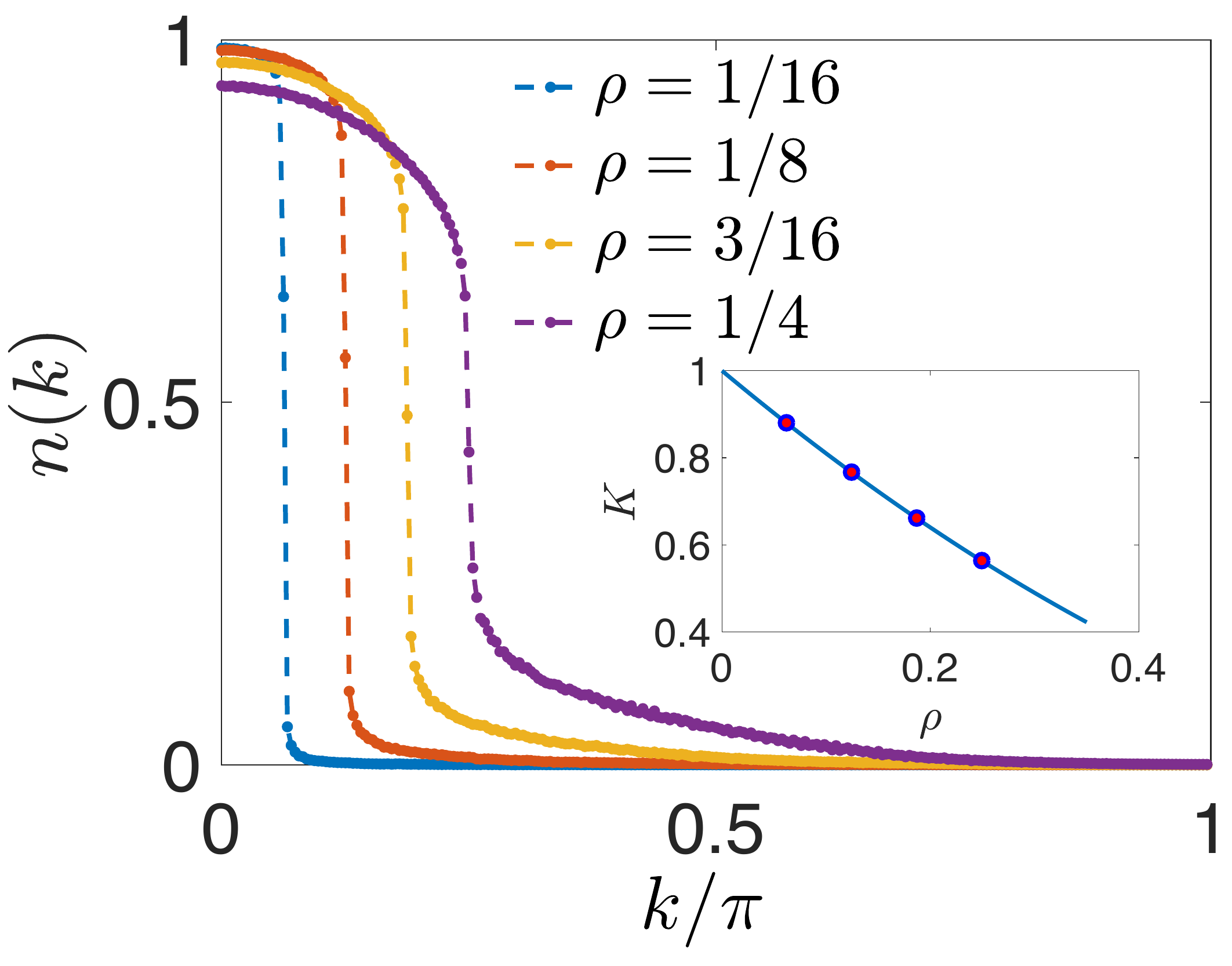}
\caption{(Color online) Momentum distribution of the ground state hard-core spinless fermions (Eq.~\eqref{eq:physicalHam}). The momentum distribution $n({\bf k})$ of the ground state is shown here, and it exhibits Luttinger liquid behavior. As we decrease the particle density, interaction effects become weaker, and the momentum distribution is approaching fermi-surface type. The interaction effects in this model can also be quantified by the deviation of the the ground state luttinger parameter $K$ from the fermion case which has $K = 1$ (see the inset). The number of lattice sites used here is $L = 512$. } 
\label{fig:interactioneffects}
\end{figure}

The algorithm developed here is also a significant technical {\it advance} in solving hard-core fermion models, which allows for calculation of much larger system-sizes than previously accessible. In Fig.~\ref{fig:interactioneffects}, we show the results for the ground state  of Hamiltonian in Eq.~\eqref{eq:physicalHam}  
which can be compared to the results in  Ref.~\onlinecite{2009_Cheong_PRB}. A straightforward variant of our algorithm could solve all finite-range interacting hard-core fermion models efficiently.

We note here that a different method for sampling slater-determinant is also developed in literature~\cite{hough2006determinantal, ScardicchioPRE}, which has been termed determinantal point process. This method also has a polynomial computational cost as ours. How the two different methods compare with each other in detail is left for a future study.

\begin{figure}
\includegraphics[width=.48\textwidth]{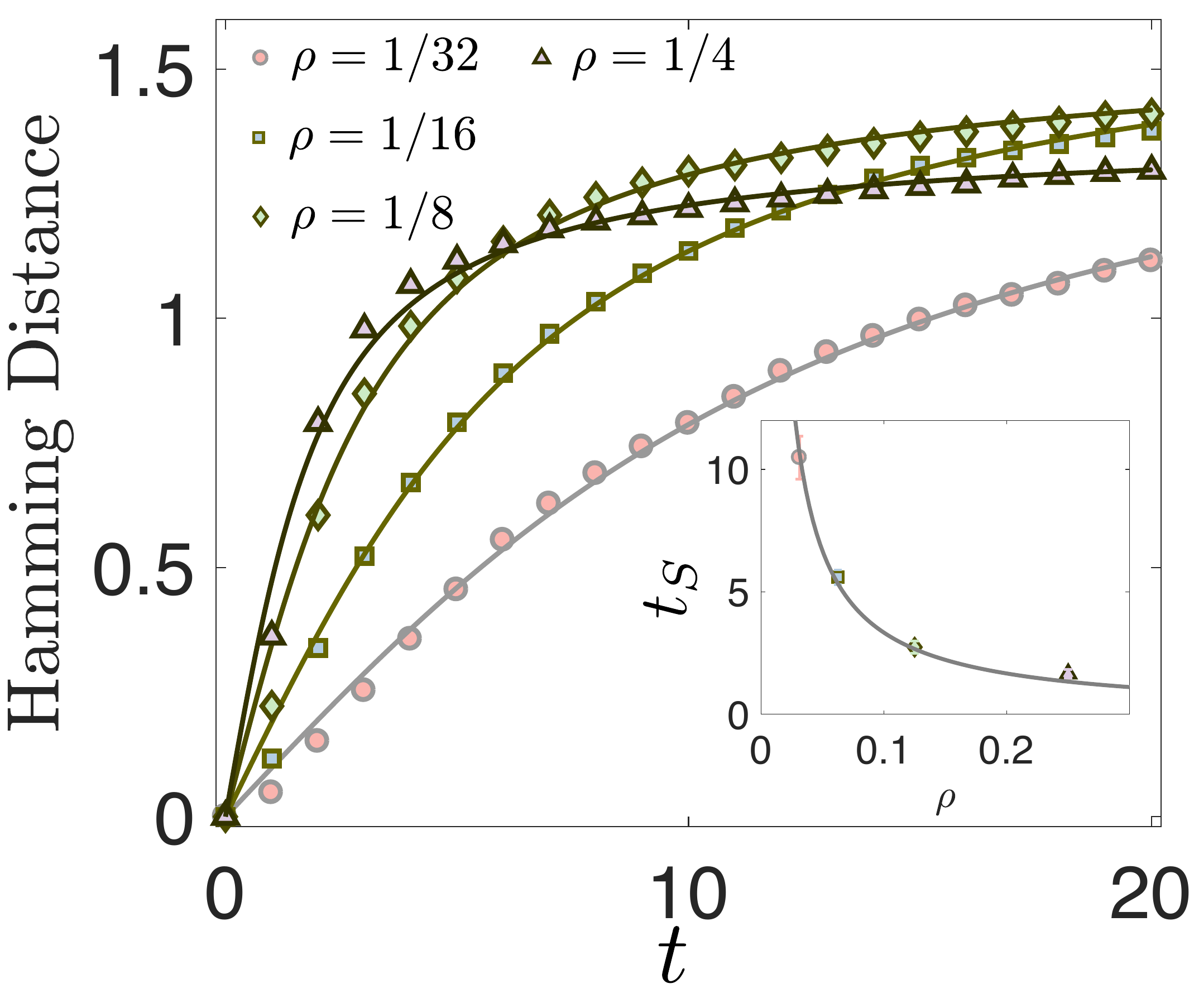}
\caption{(Color online) Quantum information scrambling in Hilbert space measured by Hamming distance. In this figure, the shown Hamming distance is normalized by $N$. The system size is fixed to be $256$ in this plot. We average over $10$ random initial states and sampling $15000$ states in Hilbert space to minimize the sampling error. The numerical results shown by symbol points fit  to a function of  ${\rm arctan} (t/t_S)$. The fitted form is shown by the lines in the plot. The characteristic time scale $t_S$ exhibits a $1/\rho$ dependence as shown in the inset. }
\label{fig:HammingDist}
\end{figure}

\section
{Hamming distance as a measure for information scrambling} 
To quantify information scrambling from few- to many-body degrees of freedom, we generalize a previously introduced Hamming distance~\cite{hauke2015many} to natural orbital occupation basis. 
%$\langle C_j^\dag C_j \rangle$. 
%The natural orbital basis $\alpha_n $ is defined from diagonalizing the time-dependent correlation function with 
%$\langle \Psi (t) | c_i ^\dag  c_{j}  | \Psi (t )  \rangle = \sum_n \lambda_n (t)  U_{i, n} (t) U_{ j, n} ^* (t) $, and $\alpha_n  = \sum_i U_{i, n} c_i$. 
In this basis, the Hilbert space is spanned by the  time-dependent states 
$
|\vec{n}; t\rangle = [C_1 ^\dag ] ^{n_1} [C_2^\dag ] ^{n_2} \ldots | 0\rangle. 
$
The Hamming distance that characterizes the distance between the dynamical state and the initial state in the Hilbert space is 
\be 
{\cal D}(t) = \sum_{\vec{n} } \left(\vec{n} - \vec{n_0}\right)^2 | \langle \vec{n}; t| \Psi(t)\rangle|^2, 
\ee 
where $\vec{n}_0$ represents the initial occupation of natural orbitals.  This Hamming distance is also a measure of {\it operator growth}~\cite{2018arXiv180202633R}. 
The natural orbital basis is defined by sorting the eigenvalues $\lambda_j$ in descending order, so that $\lambda_{j\le N} = 1$, and $\lambda_{j>N} =0$ for the initial state, and $\lambda_j \ge \lambda_{j+1}$ in general. 
For systems that do not exhibit information scrambling such as  free fermions or  many-body localized systems, ${\cal D}(t)$ remains to be an intensive quantity in dynamics, i.e., ${\cal D} (t) /L \to 0$, whereas ${\cal D} (t)/L$ becomes finite at long-time when the information spreads over the Hilbert space.

%For the dynamics controlled by Hamiltonian in Eq.~\eqref{eq:logicalHam},  $| \Psi(t) \rangle$ is a slater-determinant wave function in the logical fermion $d_j$ basis.   
A direct calculation of the Hamming distance is still exponentially difficult. We thus rewrite it in terms of observables as 
${\cal D}(t)  = 2N [1-\chi (t) ]$, with 
$
\chi (t) = \frac{1}{N} \sum_l n_{0,l} \lambda_l. 
$
Now the Hamming distance quantifying information spreading in the physical bases can be efficiently calculated at a polynomial time cost.  

The results for Hamming distance are shown in Fig.~\ref{fig:HammingDist}. The information scrambling time $t_S$ is determined by fitting the numerical results to 
$
 {\cal D}(t)  = {\cal D}_\infty {\rm arctan}(t/t_S). 
$ 
We find that at long time limit ${\cal D}_\infty$ is an extensive quantity, i.e., ${\cal D}_\infty/N$ is non-zero, and it approaches $2N(1-\rho)$ as we increases total $N$, with the fermion density $\rho = N/L$. This means the quantum information encoded in the logical $\tau$ basis scrambles to the whole many-body Hilbert space in the $\sigma$ basis. The  scrambling time $t_s$, here referring to the time for the one-dimensional  system to fully scramble, is found to have a $1/\rho$ dependence, which is qualitatively as expected because the effective interactions between physical fermions increase upon increasing the particle number density. 

\begin{figure*}
\includegraphics[width=\linewidth]{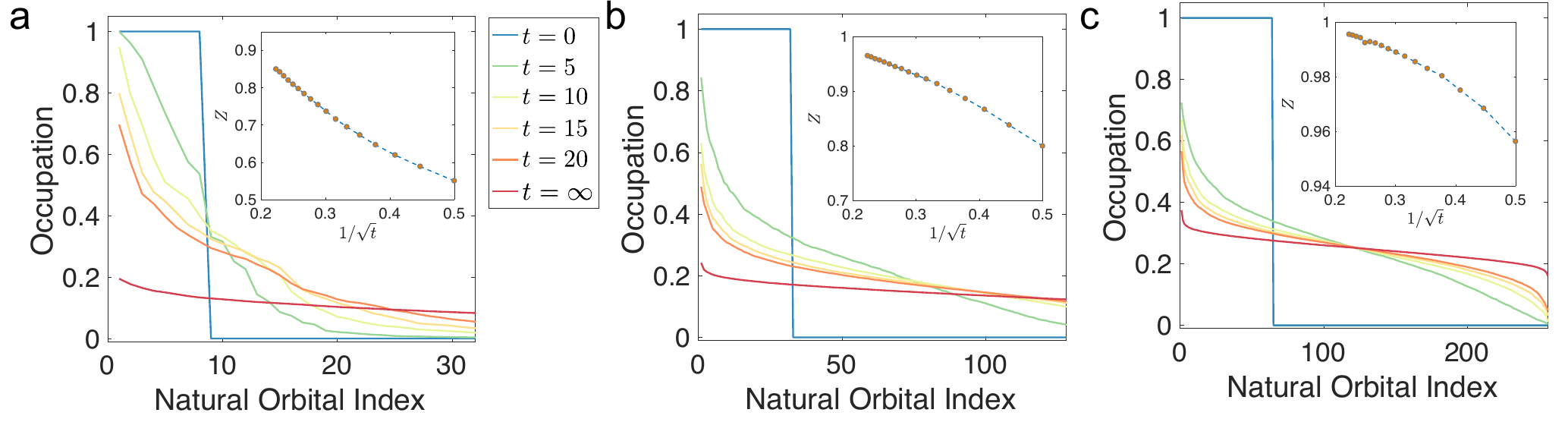}
\caption{(Color online) Correlation relaxation dynamics of the physical degrees of freedom. (a, b, c) show the occupation numbers of natural orbital basis (see main text), with different $N = 8, 32, 64$, respectively. The total system size is fixed at $L = 256$. The $t=\infty$ results correspond to $t = 10^6$ in the calculation. Despite the logical degrees of freedom $f_k$ are non-interacting, the physical ones show effective interaction effects in that natural orbital occupation exhibits relaxation behavior because otherwise they would not. The inset shows $Z(t)$ (see main text), which fits to a power law form of $1- Z(t) = \sqrt{t_r/t} +O(1/L)$. 
}
\label{fig:CorrelationRelax}
\end{figure*}

\section
{Correlation relaxation dynamics and emergence of classical diffusion at late time} 
Apart from Hamming distance, the quantum memory loss in the physical basis can also be seen in  the relaxation dynamics of correlations. Fig.~\ref{fig:CorrelationRelax} shows the evolution of the natural orbital occupation numbers, which are obtained through the eigenvalues of the correlation matrix $\langle \Psi (t) | c_i ^\dag c_j |\Psi(t)\rangle$. In the dilute limit of  $\rho \to 0$, the interaction effects are vanishing,  and the system is essentially formed by non-interacting fermions that do not relax. Away from that limit, we find efficient relaxation in the  correlation, and the relaxation rate becomes stronger as we increase the particle number $N$. To quantify the relaxation dynamics, we introduce 
$Z (t) =\frac{1}{N} \sum_l\sqrt{ \lambda_l (t) \lambda_l (t\to \infty)} $  that approaches $1$ in the long-time limit. 

We find that for large systems the long-time behavior in $Z(t)$ is captured by  
$1 - Z(t) = \sqrt{t_r/t}+O(1/L)$,   
which is attributed to classical diffusive dynamics.  
At a time $t$ much longer than microscopic time scales, the system is formed by domains of fermions in a vacuum background. 
The typical domain size is expected  to be $\sqrt{ Dt }$, with ${D}$ the diffusion constant. The late-time behavior determined by the  domain-size growth would lead to the $1/\sqrt{t}$ relaxation. This behavior also implies that a classical description could emerge from quantum many-body dynamics after the quantum information scrambles in the system. 
% a fast scrambling of quantum information of the system, because otherwise the late-time dynamics would not be described by classical diffusion physics. 

%{\it Effective model for the $1/\sqrt{t}$ relaxation dynamics.---} We devise an effective model for the observed $1/\sqrt{t}$  relaxation of natural orbital occupation. Starting from random product states, the system is formed of fermion domains on top of  a vacuum background, after certain microscopic time scales. 
The classical description consistent with numerical results is that at late time the particle density behaves as a classical superposition of localized domains, whose domain size grows according to the classical diffusion physics, i.e.,   $l_d = \sqrt{D t}$, with $D$ the diffusion constant. With the randomly distributed domains, the central limit theorem implies that the fermion density $\langle c_j ^\dag c_j \rangle$ satisfies a normal distribution with a mean $N/L$ and variance 
${\rm Var} = N/L(l_d^{-1} - L^{-1})$. The average of  $\sqrt{ \langle  c_j ^\dag c_j  \rangle }$  is then given as 
\be 
\overline{\sqrt{ \langle  c_j ^\dag c_j  \rangle }} =\sqrt{N/L} \left[ 1-\frac{1}{8} \frac{\rm Var}{(N/L)^2} \right] + O(1/L),
\ee 
from which we get $1-Z(t)\sim \sqrt{t}$, after substitution of $l_d$ by $\sqrt{D t}$.

\begin{figure*}
\includegraphics[width=\textwidth]{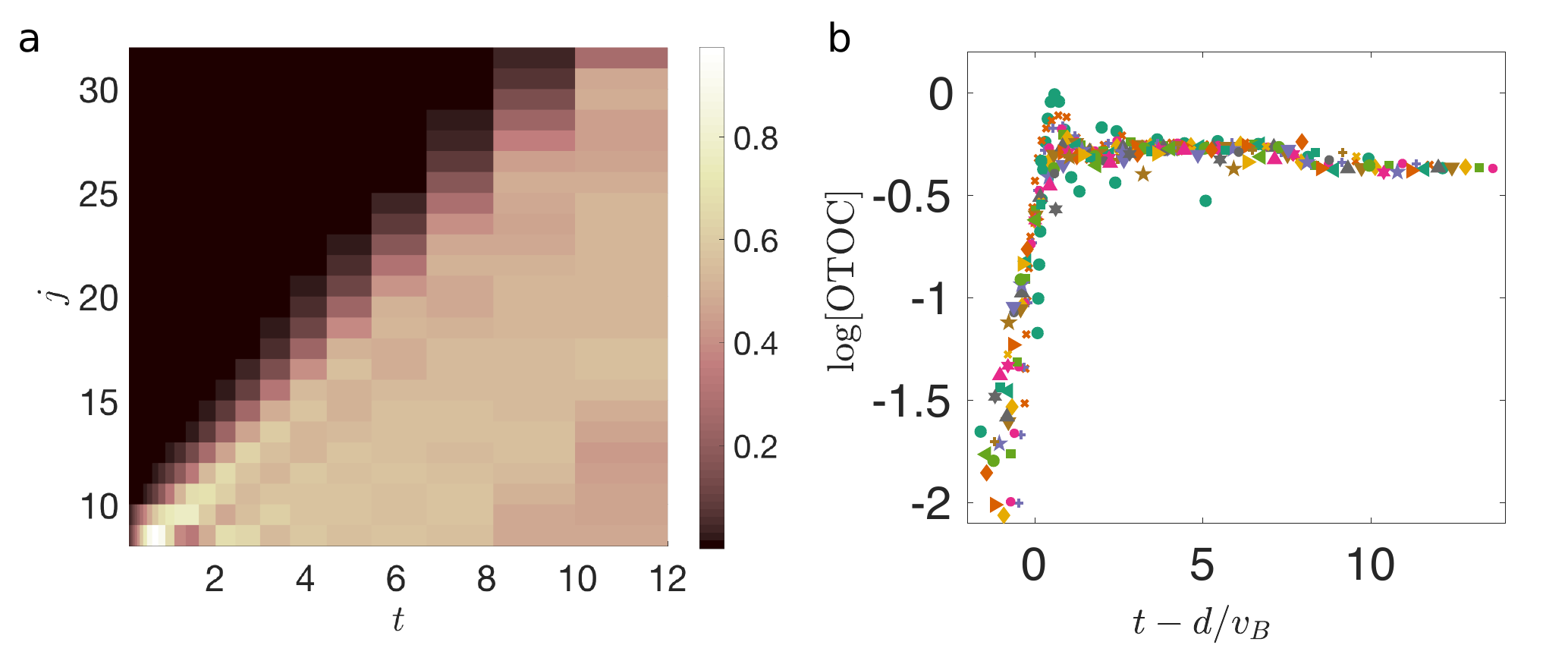}
\caption{(Color online) Out-of-time-order correlator (OTOC) of the physical operators  $G_{ij}$ in dynamics. (a) shows the color plot of OTOC of $G_{i=8, j}$. In dynamics, the correlator exhibits a light-cone behavior with a finite butterfly velocity ($v_B$). Once the operator separation range in OTOC is within the light-cone, they quickly scramble and locally equilibrate. 
In this figure,  the temperature is fixed to be $1$ here.  (b) shows the $10$-based logarithm of OTOC $G_{i=8, j} (t)$ versus $t-d/v_B$, with $d$ the operator distance $d = j-8$.    Different symbols correspond to different $j$ index. The OTOC early-time dynamics is consistent with an exponential growth. 
 In the calculation we choose system size $L = 32$ and $N = 8$.  }
\label{fig:OTOC}
\end{figure*}

\section
{Operator-mapping enabled OTOC growth}
Despite the polynomial time cost in calculating all equal-time observables, our physical system still exhibits quantum chaotic behavior in the OTOC~\cite{larkin1969quasiclassical} as in a generic thermalizing system to be explained below. The correlator we use here takes a form of 
$G_{jj'} (t) =\langle [ \sigma^z _j (t), \sigma_{j'} ^z (0) ]^\dag [\sigma^z _j  (t), \sigma^z _{j'} (0)]  \rangle_\beta$, with $\sigma_j(t)$ the Heisenberg operator, and $\langle \ldots\rangle_\beta$ the thermal ensemble average.  The sampling method we developed to calculate dynamics in few-body observables would  breakdown in calculating OTOC at long time, because the sampling error would explode as $\sqrt{2^{3V_L t} /M_s}$, with $M_s$ being the sampling number, and $v_L$ the Lieb-Robinson bound velocity. We thus evaluate OTOC with exact methods, and afford to simulate the dynamics up to a system size $L = 32$, with gpu techniques. 

Figure~\ref{fig:OTOC} shows the results for OTOC where the exponential scrambling behavior of quantum information is observed. Effective interaction effects among the physical fermions are revealed as the OTOC spreading behaves as ``ball-like" instead of ``shell-like", which differentiates interacting and free fermions~\cite{roberts2016lieb,Roberts2015}. This further confirms the information scrambling in this system. 
{The early-time behavior of OTOC gives the Lyapunov exponent $\lambda_L$ which exhibits a mild dependence of the temperature, similar to the situation in \cite{Yao:2016ayk}.} 
{Besides, the butterfly velocity is found to be approximately equal to the sound velocity in the conformal field theory at low energy, and its implication to black hole physics is worth further investigation.}

Given that the physical system exhibits quantum chaotic behavior and that the computational complexity for equal-time observables is polynomial in the system size, our quantum operator-mapping scheme is expected to motivate further researches  on the minimal microscopic computational complexity for chaotic phenomena, which is of fundamental interest in understanding complex many-body systems.

\section{Conclusion}

We have presented a novel mechanism for quantum information scrambling by hiding the information behind a highly nonlocal mapping, which relates 
the physical degrees of freedom to  the logical ones holding the quantum information.   
%This offers a novel approach to study quantum information scrambling and thermalization. 
A concrete operator mapping is provided to support this mechanism.   
We develop an algorithm to calculate physical observables, the  complexity of which is polynomial in the system size. This allows us to simulate quantum dynamics of large systems. By calculating Hamming distance, correlation relaxation and out-of-time-order correlators in dynamics, we confirm the physical system exhibits quantum information scrambling despite the fact that the mapping enabling scrambling has much lower complexity than exponential. 
Our discovery offers a better understanding of computation complexity, quantum thermalization, and information scrambling, and the connections among them.  
Furthermore our approach can be  readily generalized to construct exact solutions for novel interacting quantum dynamics. Building nontrivial but tractable mappings from quantum integrable models may inspire ideas to formulate quantum Kolmogorov-Arnold-Moser theorem~\cite{kolmogorov1979preservation,fioretto2010quantum}, an important open question relevant to a broad range of physics spanning quantum chaos and quantum many-body localization.

\medskip
\ncsection{Acknowledgement}

We acknowledge helpful discussion with Zhexuan Gong, Yuan-Ming Lu, Meng Cheng, Lei Wang, and Sriram Ganeshan.  This work is supported by National Program on Key Basic Research Project of China under Grant No. 2017YFA0304204 (XL), National Natural Science Foundation of China under Grants No. 117740067 (XL),  and the Thousand-Youth-Talent Program of China (XL).  
G.Z. is supported by US ARO-MURI and US YIP-ONR. 
M.H. acknowledges support from the US National Science Foundation through grant PHY-1602867, and the Start-up Grant at Florida Atlantic University, USA.
X.W. acknowledges support from the Research Grants Council
of the Hong Kong Special Administrative Region, China
(No. CityU 11303617), the National Natural Science
Foundation of China (No. 11604277), and the Guangdong
Innovative and Entrepreneurial Research Team Program
(No. 2016ZT06D348).
This work was completed at the Aspen Center for Physics, which is supported by US National Science Foundation Grant PHY-1607611.

\bibliography{references}

\end{document}